%% file: BurringtonJardinePeet-fractionalLmodes-Version2.tex
\documentclass[12pt]{article}
\usepackage{amsmath}
\usepackage{amssymb}
\usepackage{graphicx}
\usepackage{hyperref}
\usepackage{enumitem}
\usepackage{color}

\begin{document}

\numberwithin{equation}{section}

\begin{titlepage}
\hbox to \hsize{\hspace*{0 cm}\hbox{\tt }\hss
   \hbox{\small{\tt }}}

\vspace{1 cm}

\centerline{\bf \Large The OPE of bare twist operators in}
\medskip
\centerline{ \bf \Large bosonic $S_N$  orbifold CFTs at large $N$}

\vspace{1 cm}

\vspace{1 cm}
\centerline{\large Benjamin A. Burrington$^{\star}$\footnote{benjamin.a.burrington@hofstra.edu},
Ian T. Jardine$^\dagger$\footnote{jardinei@physics.utoronto.ca}\,, and
Amanda W. Peet$^{\dagger\S}$\footnote{awpeet@physics.utoronto.ca}}

\vspace{0.5cm}

\centerline{\it ${}^\star\!\!$ Department of Physics and Astronomy, Hofstra University, Hempstead, NY 11549, USA}
\centerline{\it ${}^\dagger$Department of Physics, University of Toronto, Toronto, ON M5S 1A7, Canada}
\centerline{\it ${}^\S$Department of Mathematics, University of Toronto, Toronto, ON M5S 2E4, Canada}

\vspace{0.3 cm}

\begin{abstract}

In this work, we explore the twist operator OPEs of a generic bosonic symmetric product ($S_N$) orbifold CFT. We conjecture that at large $N$ the OPE of bare twist operators contains only bare twists and excitations of bare twists with fractional Virasoro modes. These fractionally excited operators are the only ones that depend exclusively on the lengths of the twists and the central charge, agreeing with the general structure of correlators of bare twists found in the literature. To provide evidence for this, we study the coincidence limit of a four point function of bare twist operators to several non-leading orders. We show how the coefficients of these powers can be reproduced by considering bare twist operators excited by fractional Virasoro modes in the exchange channels.

\end{abstract}

\end{titlepage}

\tableofcontents

\section{Introduction}

In the twenty years since Maldacena first introduced holography \cite{Maldacena:1997re}, there has been significant progress made in understanding this duality between gravity and conformal field theories. One major line of investigation has been to determine what properties a CFT must possess in order to have a holographic dual. Early results \cite{Heemskerk:2009pn} showed that having an Einstein gravity dual requires a large central charge and constrains the CFT to have a sparse spectrum of low-lying operators. These conditions are necessary, but it is still an open question whether they are sufficient. For bulk duals not well approximated by Einstein gravity, other conditions may be expected. Still, it is an important goal to find CFTs that satisfy these basic requirements.

One set of CFTs that has been explored for their holographic properties are permutation orbifold CFTs \cite{Keller:2011xi,Haehl:2014yla,Belin:2014fna,Benjamin:2015hsa,Belin:2015hwa,Benjamin:2015vkc,Belin:2016yll,Belin:2017jli}. These theories have a built-in way of satisfying the two restrictions. Having large numbers of copies gives the required large central charge, but at the cost of a large number of low lying states. The orbifold then restricts states, giving a more sparse result.

A well studied subset of these orbifold CFTs is symmetric permutation orbifolds. The conditions these satisfy imply that they are dual to strongly quantum gravity, not semiclassical gravity, as the growth of states in the high energy regime is Hagedorn. The reason for this growth is that although the action of the orbifold projects out many states for a copied CFT, it adds in new states with twisted boundary conditions. These twisted sector states dominate the density of states of the theory at high energy. Understanding the twisted sector is important to understanding the general structure of orbifold CFTs.

One specific symmetric orbifold CFT that has been explicitly shown to play a role in holography, and in microscopic modelling of black hole physics, is the D1D5 CFT at the orbifold point. At this specific point in moduli space, the CFT is a 1+1 dimensional symmetric permutation orbifold. For a review of the D1D5 system, see \cite{David:2002wn}. In order to recover a dual description that is semiclassical, one has to deform the theory using an operator from the twisted sector.  This gives extra incentive for understanding twist operators, beyond their dominance at high energy. There has been a large body of work studying the effects of twist operators on states in this CFT \cite{Avery:2010hs,Carson:2014yxa,Carson:2014xwa,Carson:2014ena,Burrington:2014yia,Carson:2015ohj,Carson:2016cjj,Carson:2016uwf,Carson:2017byr}. 

Our work here will take a slightly different approach. In an earlier work \cite{Burrington:2017jhh}, we used the techniques of \cite{Lunin:2000yv,Lunin:2001pw,Burrington:2012yn} to lift computations of an $S_N$ orbifold CFT to a covering space, where the boundary conditions associated with twist operators are encoded in the lifting map.  In the lift, operators on the base are lifted to corresponding operators in the cover, and the computation is dressed with a Liouville term accounting for the conformal anomaly, which depends on the central charge $c$ of the base CFT (copied $N$ times to construct the orbifold). In \cite{Burrington:2017jhh} we found that the OPE of operators on the base could be read off from the covering space OPE, but only in the restricted case that one of the operators was in the untwisted sector.  Taking OPEs by moving twist sector operators seemed to be less straightforward to analyze, because of two complications: the map to the cover changes, and this in turn alters the Liouville term.  While \cite{Burrington:2017jhh} dealt with the specific D1D5 CFT near the orbifold point, we expect that the difficulties of moving twist operators to find their OPEs will persist for all orbifold CFTs, because the need to interpret the change in the map from the base to the cover and resulting changes to the Liouville term is always present.  For this reason, we start by working with the simplest class of twist operators available: the bare twists of a bosonic $S_N$ orbifold CFT, the topic of this paper.

The bare twists of an orbifold CFT are the lowest conformal dimension operators in a given twist sector.  The general structure of the correlators of bare twists for bosonic $S_N$ orbifold CFTs was explored in \cite{Lunin:2000yv}.  First, the bare twists are lifted to operator insertions of $1$ in the covering surface, making the lifted computation trivial, pushing all of the information about the correlator into the Liouville action which depends on the map to the cover.  Further, in \cite{Lunin:2000yv}, this general structure of $n$-point functions of bare twists showed that the large $N$ behavior was universal, depending only on the length of the twists (the length of the cycles in $S_N$), the central charge of the base CFT ($c$), and the total number of copies of the CFT ($N$) appearing in the orbifold.  This raises the question of what type of operators could appear in the OPE of two such operators, given that the four point functions contain information about the OPE through the crossing channels. 

In fact, this structure gives some strong clues:-
\begin{itemize}[noitemsep,topsep=3pt]
\item The OPE between two bare twists should produce a set of operators that, once lifted, give a non trivial contribution when accompanied by insertions of 1.  Thus, the OPE should contain operators that once lifted contain c-number operators.
\item These c-number operators should only depend on the same information that the correlators do: the lengths of twists, the central charge of the base CFT, and the total number of copies.
\item These operators should be available for all $S_N$ orbifold CFTs, such that the universal nature of the large $N$ behavior of correlators can be reproduced.
\end{itemize}

These clues seem to be very limiting, and lead us to conjecture that the OPE of twist operators will consist of only bare twist operators and bare twist operators excited by fractional modes of the stress energy tensor $T$, which we will also refer to as fractional Virasoro modes.  These candidate modes meet our criteria outlined in the last paragraph.  First, they are clearly always available for every $S_N$ orbifold CFT, since each individual copy of the CFT contains a copy of the stress tensor, and the fractional modes are constructed using these individual copies (this is explained more thoroughly in the next section).  Next, the lift to the covering surface does contain c-number parts to the operator, owing to the non-tensorial transformation of the stress tensor under conformal maps, i.e., the Schwarzian derivative term.  Further, this Schwarzian derivative term will only have information about the central charge, and the map, which only contains information about the size of the cycles.  Hence, these modes are natural candidates to build the OPE of two bare twists.

To provide evidence for this conjecture, we will examine a four point function of bare twist operators and work out the expansion of this correlator in the coincidence limit. With this information, we will show how the twist operators and their excitations can fully account for the exchange channels in the four point function for a number of powers of the expansion.  Further, continuing this logic, the OPE between any two of these operators should also only depend on the three criteria above, for consistency.  Hence, we expect that the bare twists excited by fractional modes of the stress tensor form a universal set of operators, available in any bosonic $S_N$ orbifold CFT, that are closed under the operator algebra at large $N$.  

The layout of the paper is as follows. In section \ref{fracmodesec}, we will give a short review of symmetric orbifold CFTs and twist operators. We will then discuss how we can use the lifting technique of \cite{Lunin:2000yv,Lunin:2001pw} to generalize the Virasoro algebra to fractional modes. We will use this algebra to study operators created using fractional Virasoro modes, showing that we can create an infinite number of new primaries. In section \ref{coinclimit}, we will review the computation of the four point function $\langle \sigma_n\sigma_2\sigma_2\sigma_n\rangle$ of two length-2 twists with two length-$n$ twists, 
and show how we can find the coincidence limit order by order. In section \ref{exchangereconstruct}, we will reconstruct the coefficients of this limit by using  operators created via fractional Virasoro modes and the bare twist operator. We will finish with a discussion in section \ref{discussion}.

\textbf{Note}: Shortly after submission of this preprint, we noticed \cite{Roumpedakis:2018tdb}, which has overlap with our work.

\section{Fractional modes of the stress energy tensor}\label{fracmodesec}
\subsection{Symmetric orbifold CFT and the twisted sector}\label{SNOrbifold}

Our focus for this paper will be on general symmetric orbifold CFTs, denoted by $M^N/S_N$. Here  $M$ stands for the seed CFT of which there are $N$ copies. If this seed CFT has a field $\phi$, then we will denote the $\phi$ field of the $i$th copy of the seed CFT as $\phi_{(i)}$ where $i$ can run from 1 to $N$. This implies that the central charge of the orbifold theory is
\begin{equation}
c_{tot}=c\,N\,,
\end{equation}
where $c$ is the central charge of the seed CFT. We will take $N$ to be large throughout.

Operators in this orbifold CFT are created using these $N$ copies, but the final result must be $S_N$ invariant. Consider first the untwisted sector. Here an operator is given by simple sums and products over the copies to give an $S_N$ invariant result. The more interesting sector is the twisted sector.  The states on the cylinder of the twisted sector obey boundary conditions such that fields are periodic up to the action of the orbifold symmetry, in our case $S_N$.  The operators that are dual to these states are the twist operators, and the twist operator of lowest conformal weight in a given twist sector is called the bare twist.  Further, since the group $S_N$ is constructed using products of cycles, understanding the operators associated with cycles is sufficient to understand the $S_N$ orbifold.  As an example, consider an operator $\sigma_{(1...n)}(0)$ which implements the boundary condition
\begin{equation}
\phi_{(1)}\rightarrow\phi_{(2)}\rightarrow...\rightarrow\phi_{(n)} \,,
\end{equation}
as we go around $z\rightarrow z\,e^{2\pi i}$. This means we can identify a bare twist operator associated with a cycle simply by its length. In our example we say that it is a twist $n$ operator and we will often shorten the notation to $\sigma_n$. The conformal weight of the bare twist depends on its length $n$ through the formula
\begin{equation}
h=\frac{c}{24}\left(n-\frac{1}{n}\right)\,.
\end{equation}
However, the example operator on its own is not $S_N$ invariant, as it picks out the first $n$ operators. A fully $S_N$ twist operator is given by a normalized sum over the full orbit of a particular representative permutation. The $S_N$ invariant combination is given by \cite{Lunin:2000yv,Lunin:2001pw}
\begin{equation}
\frac{1}{\sqrt{N!\,n(N-n)!}}\sum\limits_g\sigma_{g(12..n)g^{-1}}\,,
\end{equation}
where $g$ are the group elements of $S_N$.

We can consider excitations of the bare twist to fill out the twisted sector. These twist operators give access to fractionally moded excitations, which are not accessible in the untwisted sector, and generally these will depend on the details of the seed CFT. These are written in the form
\begin{equation}\label{fractionalmoding}
\phi_{-m/n}=\oint\frac{dz}{2\pi i}\sum\limits_{k=1}^{n}\phi_{(k)}(z)e^{-2\pi i m (k-1)/n}z^{h-m/n-1}\,.
\end{equation}
where $h$ is the weight of the field $\phi$. We can see that these fractional modes only make sense in the presence of a twisted operator. Taking $z\rightarrow ze^{2\pi i}$ will give a phase  from the fractional power of $z$ which can only be canceled by permuting the copies inside of the sum. These fractional modes are key in symmetric orbifolds with supersymmetry, as fractionally moded R-symmetry currents are used to build up the chiral primaries of the twisted sector.

What will interest us in this paper are the fractional modes of the Virasoro operator
\begin{equation}
L_{-m/n}=\oint\frac{dz}{2\pi i}\sum\limits_{k=1}^{n}T_{(k)}(z)\, e^{-2\pi i m (k-1)/n}\,z^{1-m/n}\,.
\end{equation}
Here we use a slightly loose notation: note that the copies of the stress tensor involved are only those parallel to the cycle (the first $n$ indices, in the formula above).  When dealing with the full stress tensor of the theory, all copies would need to be included.  However, from context, the meaning of such expressions will be clear.

These particular operators are available for any seed CFT chosen, and thus these fractional modes will be universal to any symmetric orbifold CFT. We expect that the only operators present in bare twist OPEs are bare twists and excitations by these fractional modes. We will provide evidence for this conjecture by exploring the $\sigma_n\times\sigma_2$ OPE.  Continuing this line of reasoning, we further conjecture that the $S_N$ orbifold contains a universal subsector, spanned by the bare twists and their fractional Virasoro descendants, and that these operators have a closed algebra at leading order in large $N$.

To check this, we will need some way to compute OPEs and correlation functions with twisted boundary conditions. The main method we will use to examine these fractionally moded Virasoro modes will be the Lunin-Mathur technique \cite{Lunin:2000yv,Lunin:2001pw}. The main idea of the technique is to lift twisted operators from the base space to a covering space where the twisted boundary conditions are ramified. This lift is done using some map between the base, denoted with the $z$ coordinate, and the cover, denoted with the $t$ coordinate. This map has ramified points $t_i$ which implement the boundary condition for a twist operator inserted at $z_i$. These will take the form 
\begin{equation}\label{generalmap}
z-z_i=(t-t_i)^n\left(c_0+c_1(t-t_i)+c_2(t-t_i)^2+\cdots\right)\,.
\end{equation}
The bare twists are then represented by insertions of the identity on the cover. Their contribution to correlation functions are given by the exponential of the action of the Liouville field that keeps track of Weyl anomaly due to the non-zero (large) central charge.

This will be key to our work later. On the cover, the bare twists are identities, so whatever operator appears in their OPE must also have some piece proportional to the identity. When we lift to the cover, the stress energy tensor does not transform as a full primary. The extra piece, the Schwarzian of the lifting map, is proportional to the identity and will give the contribution of the stress energy tensor to the correlator. The Schwarzian term depends only on $c$ and the map itself, which is fixed fully by the ramifications given by the twist insertions. So we can see that, when viewing the OPE from the cover, the fact that the fractional Virasoro modes contribute should be obvious.

\subsection{Fractional Virasoro algebra}

Before we dive into giving evidence for our conjecture, we can first exhibit some properties of these fractional Virasoro modes. We will start by showing how to extend the usual commutation relation for Virasoro modes to fractional modes. We lift the following combination of Virasoro modes acting on a general twisted operator to the covering space,
\begin{align}
& \left[L_{k/n}, L_{k'/n}\right]| \sigma' \rangle \rightarrow \nonumber \\
& \oint \frac{dt_2}{2\pi i}\oint_{t_1=t_2} \frac{dt_1}{2\pi i} z(t_1)^{k/n+1}z(t_2)^{k'/n+1}\left(\frac{dz}{dt_1}\right)^{-1}\left(\frac{dz}{dt_2}\right)^{-1} \nonumber \\
& \left(T(t_1)-\frac{c}{12}\left\{z(t_1),t_1\right\}\right) \left(T(t_2)-\frac{c}{12}\left\{z(t_2),t_2\right\}\right) \sigma'_{\uparrow}\,.
\end{align}
where we have denoted the twist operator on the cover as $\sigma'_{\uparrow}$. The contour integral of $t_1$ around $t_2$ just picks out the simple pole as $t_1$ approaches $t_2$.  Note that since $t_2$ is a non-ramified point, the Schwarzian remains finite, as does the term $z(t_1)^{k/n+1}({dz}/{dt_1})^{-1}$, and so the only singularities come from $T-T$ OPEs on the cover.  Thus,
\begin{align}
& \left[L_{k/n}, L_{k'/n}\right]| \sigma' \rangle \rightarrow \\
& \oint \frac{dt_2}{2\pi i}\oint_{t_1=t_2} \frac{dt_1}{2\pi i} z(t_1)^{k/n+1}z(t_2)^{k'/n+1}\left(\frac{dz}{dt_1}\right)^{-1}\left(\frac{dz}{dt_2}\right)^{-1}\left(\frac{c}{2t_{12}^4}+\frac{2T(t_2)}{t_{12}^2}+\frac{\partial T(t_2)}{t_{12}}\right) \sigma'_{\uparrow}\,.\nonumber
\end{align}
where we have dropped the finite terms because they do not contribute to the contour integral of $t_1$ near $t_2$.

We can evaluate the $t_1$ integral using our usual residue integral identities, to find
\begin{align}
& \left[L_{k/n}, L_{k'/n}\right]| \sigma' \rangle \rightarrow \\
& \oint \frac{dt_2}{2\pi i}z(t_2)^{k'/n+1}\left(\frac{dz}{dt_2}\right)^{-1}
\left(\partial T(t_2)z(t_2)^{k/n+1}\left(\frac{dz}{dt_2}\right)^{-1}+  2T(t_2)\frac{d}{dt_2}\left(z(t_2)^{k/n+1}\left(\frac{dz}{dt_2}\right)^{-1}\right) \right. 
\nonumber\\
&\left. \qquad\qquad\qquad \qquad\qquad\qquad +\frac{c}{2}\frac{1}{3!}\frac{d^3}{dt_2^3}\left(z(t_2)^{k/n+1}\left(\frac{dz}{dt_2}\right)^{-1}\right)\right) \sigma'_{\uparrow}\,. \nonumber
\end{align}
The integrand may be freely integrated by parts, since the contour is closed, and this allows us to combine the first two terms involving $T$.  This gives
\begin{align}
& \left[L_{k/n}, L_{k'/n}\right]| \sigma' \rangle \rightarrow \\
& \oint \frac{dt_2}{2\pi i}\left(\frac{dz}{dt_2}\right)^{-1}
\!\!
\left(z(t_2)^{\frac{k+k'}{n}+1}\frac{(k-k')}{n} T(t_2)+\frac{c}{12}z(t_2)^{k'/n+1}\frac{d^3}{dt_2^3}\left(z(t_2)^{k/n+1}\left(\frac{dz}{dt_2}\right)^{-1}\right)\right)\sigma'_{\uparrow}\,.\nonumber
\end{align}
The first term almost looks like a fractional mode of the stress tensor on the cover, although it is missing the Schwarzian.  We introduce this by adding and subtracting it, writing
\begin{align}
& \left[L_{k/n}, L_{k'/n}\right]| \sigma' \rangle \rightarrow \\
& \oint \frac{dt_2}{2\pi i} \left(\frac{dz}{dt_2}\right)^{-1}\left( z(t_2)^{(k+k')/n+1}\frac{(k-k')}{n} \left(T(t_2)-\frac{c}{12}\left\{z(t_2),t_2\right\}\right)
\right. \nonumber \\
& \left.+\frac{c}{12}\left(z(t_2)^{k'/n+1}\frac{d^3}{dt_2^3}\left(z(t_2)^{k/n+1}\left(\frac{dz}{dt_2}\right)^{-1}\right)+\frac{(k-k')}{n}z(t_2)^{(k+k')/n+1}\left\{z(t_2),t_2\right\}\right)\right)\nonumber\sigma'_{\uparrow}\,.
\end{align}
All terms above are manifestly antisymmetric under the interchange of $k\leftrightarrow k'$ except the triple derivative term in the second line.  We may remedy this by realizing that we may integrate by parts. Once integrated by parts, this term gives
\begin{equation}
\oint \frac{dt_2}{2\pi i}\frac{c}{12}\frac{(k-k')(k^2+kk'+(k')^2-n^2)z(t_2)^{(k+k')/n-1}}{2n^3} \left(\frac{dz}{dt_2}\right)\,,
\end{equation}
and so is a total derivative and gives zero unless $k+k'=0$.

In the case $k+k'=0$, we see that this simplifies to
\begin{equation}
\oint \frac{dt_2}{2\pi i}\frac{c}{12}\left(\left(\frac{k}{n}\right)^2-1\right)\frac{k}{n}z(t_2)^{-1}\left(\frac{dz}{dt_2}\right)\,.
\end{equation}
For a map of the form in equation (\ref{generalmap}), the simple pole is given by $n/t$, and so
\begin{equation}
\oint \frac{dt_2}{2\pi i}\frac{c}{12}\left(\left(\frac{k}{n}\right)^2-1\right)\frac{k}{n}z(t_2)^{-1}\left(\frac{dz}{dt_2}\right)=\frac{cn}{12}\left(\left(\frac{k}{n}\right)^2-1\right)\frac{k}{n}\,.
\end{equation}
Thus, in total, we find that
\begin{align}\nonumber
\left[L_{k/n}, L_{k'/n}\right]| \sigma' \rangle\rightarrow &\oint \frac{dt_2}{2\pi i} \left(\frac{(k-k')}{n}z(t_2)^{\frac{k+k'}{n}+1}\left(\frac{dz}{dt_2}\right)^{-1} \left(T(t_2)-\frac{c}{12}\left\{z(t_2),t_2\right\}\right)\sigma'_{\uparrow}\right)\\
&+\delta_{k+k',0}\,\frac{cn}{12}\left(\left(\frac{k}{n}\right)^2-1\right)\frac{k}{n}\,\sigma'_{\uparrow}\,.
\end{align}
The first term is just a fractional mode of the stress tensor on the cover, and so we may interpret the above expression on the base space as
\begin{equation}
\left[L_{k/n}, L_{k'/n}\right]| \sigma' \rangle=\left(\frac{k-k'}{n}L_{(k+k')/n}+\delta_{k+k',0}\,\frac{cn}{12}\left(\left(\frac{k}{n}\right)^2-1\right)\frac{k}{n} \right) | \sigma' \rangle\,.
\end{equation}
Thus, the algebra of fractional modes of the stress tensor is
\begin{equation}\label{fractionalalgebra}
\left[L_{k/n}, L_{k'/n}\right]=\frac{k-k'}{n}L_{(k+k')/n}+\delta_{k+k',0}\,\frac{cn}{12}\left(\left(\frac{k}{n}\right)^2-1\right)\frac{k}{n}\,,
\end{equation}
which is just the Virasoro algebra, generalizing the subscripts to be fractions (note that the central term refers to $n\times c$, i.e., only to the copies parallel to the cycle, as mentioned before).  This algebra is independent of the details of the map, which is not a priori obvious for the fractional modes. This identity will prove useful in the upcoming subsection for computing the norm of states and determining whether operators created with these fractionally moded Virasoro operators are primary.

\subsection{Primary and non-primary fractional Virasoro operators}\label{primaryfractionalmodes}

In this subsection, we will characterize operators of the form $L_{-k/n}\sigma_n$ with $k>0$. We can, of course, create operators with multiple products of the fractional modes, but we will focus on single excitations for simplicity. Obviously if $k\propto n$, then this is not a fractional mode excitation but a descendant of the bare twist $\sigma_n$ (ignoring the copies perpendicular to the cycle). When this is not the case, we will show that operators with fractional modes are primary operators or have some primary component to them.

First, we can consider the lowest lying excitation
\begin{equation}
L_{-1/n}\sigma_n.
\end{equation}
We can show by lifting this to the cover that this operator is actually zero for all $n$. Consider the lifting with a general map like in equation (\ref{generalmap}), which gives
\begin{align}
&L_{-1/n}\sigma_{n}\rightarrow \oint \frac{dt}{2\pi i}z^{1-1/n}\left(\frac{dz}{dt}\right)^{-1}\left(T(t)-\frac{c}{12}\{z,t\}\right)\\
&= \oint \frac{dt}{2\pi i}\frac{(c_0+c_1t+\cdots)^{1-1/n}}{(c_0 n+ c_1(n+1)t+\cdots)} \left(T(t)-\frac{c}{12}
\left(\frac{1-n^2}{2t^2}-\frac{c_1(n^2-1)}{nc_0 t}+\cdots\right)\right)\,.
\end{align}
It is immediately obvious that the first term vanishes, but the Schwarzian does have some non-trivial singularities that could contribute. Looking more closely yields
\begin{align}
&\frac{(n^2-1)c}{12}\left(\frac{1}{2}\partial\left[\frac{(c_0+c_1t+\cdots)^{1-1/n}}{(c_0 n+ c_1(n+1)t+\cdots)}\right]_{t=0}+\frac{c_1}{nc_0}\frac{c_0^{-1/n}}{n}\right)\\
&=\frac{(n^2-1)c}{12}\left(\frac{(n-1)c_0^{1-1/n}c_1-c_0^{1-1/n}c_1(n+1)}{c_0n^2}+\frac{c_1c_0^{1-1/n}}{n^2c_0^2}\right)\\
&=0\,.
\end{align}
This is an interesting result, as it is not obvious from the base space why this operator is zero. This result comes directly from considering the ramification around $z=0$ and lifting.  In some ways, this may be a more algebraic way of defining what we mean by the bare twist operator: it is the operator such that $L_{-1/n}| \sigma_n \rangle=0$ and $L_{a/n}| \sigma_n \rangle=0$ for $a>0$.

We can double check this using the fractional Virasoro algebra we found in equation (\ref{fractionalalgebra}) to work out the normalization of the state. For $k>0$, our state $L_{-k/n}\sigma_n$ has the norm given by
\begin{equation}\label{normformkdn}
\langle L_{-k/n}\sigma_{n}|L_{-k/n}\sigma_{n}\rangle=\frac{ck(k^2-1)}{12n^2}\,.
\end{equation}
We can see that if we plug in $k=1$, we get zero which shows it is indeed a null state. However, this is clearly non-zero for $k>1$, so we may have some interesting states to examine.

If we want to show that $L_{-k/n}\sigma_n$ is a conformal primary, then we must show that $L_aL_{-k/n}|\sigma_n\rangle$ vanishes for all $a>0$ (with $a$ integer). With the fractionally moded algebra, this is straightforward
\begin{equation}
L_{a} L_{-k/n} |\sigma_{n} \rangle=[L_{an/n}, L_{-k/n}] |\sigma_{n} \rangle=
\left(a+{\frac{k}{n}}\right)L_{a-k/n}|\sigma_{n}\rangle\,.
\end{equation}
This will vanish if $a-k/n\geq-1/n$, as any positive mode will cause the primary $\sigma_n$ to vanish and we have shown that $L_{-1/n}\,\sigma_n$ vanishes as well. So then we have the condition $k\leq an+1$. If this is true for all $a$, then the strongest bound is then when $a=1$, of $k\leq n+1$.

Even though the operators with $k>n+1$ are not primaries, they can be combined with descendants to make new primaries. Consider the operator $L_{-(n+2)/n}\sigma_{n}$. We can determine if it is quasi-primary by the action of $L_1$ ,
\begin{equation}
L_1L_{-(n+2)/n}\,\sigma_{n}=L_{-(n+2)/n}L_1\,\sigma_{n}+\frac{2n+2}{n}L_{-2/n}\,\sigma_{n}=\frac{2n+2}{n}L_{-2/n}\,\sigma_{n}\,,
\end{equation}
which is not zero. So $L_{-(n+2)/n}\,\sigma_{n}$ is not a quasi-primary operator by itself, but we can consider taking a linear combination with $L_{-1}L_{-2/n}\sigma_n$. This is equivalent to doing a projection procedure discussed in Appendix A of \cite{Burrington:2017jhh}. This will give the quasi-primary part of the operator above,
\begin{align}\label{Lmn3dn1}
&L_{-(n+2)/n}\,\sigma_{n}-\frac{1}{2h-2}L_{-1}L_1L_{-(n+2)/n}\,\sigma_{n}\\\nonumber
&=L_{-(n+2)/n}\,\sigma_{n}-\frac{24(n+1)}{c(n^2-1)+48}L_{-1}L_{-2/n}\,\sigma_{n}\,.
\end{align}
We can generalize this procedure to any $k>n+1$, giving an infinite number of primary operators using these fractionally moded Virasoro modes. In the next section, we will show how the primary fractionally moded Virasoro operators can fully account for the exchanges in a four point function.

\section{Coincidence limit for bare twists}\label{coinclimit}
\subsection{Four point function from Lunin-Mathur}

In order to test our conjecture, we will consider the coincidence limit of the normalized four point function
\begin{equation}
\frac{\langle\sigma_n(0,0)\sigma_2(1,1)\sigma_2(w,\bar{w})\sigma_n(\infty,\bar{\infty})\rangle}{\langle\sigma_2(0,0)\sigma_2(1,1)\rangle\langle\sigma_n(0,0)\sigma_n(\infty,\bar{\infty})\rangle}\,.
\end{equation}
This is will tell us about the $\sigma_2\times\sigma_n$ OPE, not a general $\sigma_m\times\sigma_n$ OPE, but we expect the general form of the OPE to be the same. The details of the computation are given in \cite{Arutyunov:1997gt}\cite{Arutyunov:1998eq}\cite{Lunin:2000yv}, but we will review some of the important aspects here.

First, we should analyze the overlap of the indices in the $\sigma_2\times\sigma_n$ OPE. There are three options: zero, one or two overlaps. In the first case, the correlator will just factorize, i.e., the OPE is trivial. In the second case, the two twist operators will fuse into an operator with $\sigma_{n+1}$. In the third case, the OPE will result in a $\sigma_{n-1}$ operator. We can compute the genus of the covering surface for both these cases using the Riemann-Hurwitz formula
\begin{equation}
g=\frac{1}{2}\sum\limits_jr_j-s+1\,,
\end{equation}
where $r_j$ are the ramifications for each of the twist operators and $s$ is the total number of sheets involved. Note that $r_j=n_j-1$, where $n_j$ is the length of the twist. For the one overlap case, we find
\begin{equation}
g=\frac{1}{2}+\frac{1}{2}+\frac{n-1}{2}+\frac{n-1}{2}-(n+1)+1=0\,,
\end{equation}
and for the two overlap case we find
\begin{equation}
g=\frac{1}{2}+\frac{1}{2}+\frac{n-1}{2}+\frac{n-1}{2}-n+1=1\,.
\end{equation}
As discussed in section \ref{SNOrbifold}, an $S_N$ invariant correlator will involve a sum of the disconnected, genus zero, and genus one pieces as we sum over the different permutations in the orbit. Since the disconnected piece will not contain information about $\sigma_2\times\sigma_n$, we can ignore it. Also, we do not expect that the operators showing up in the OPE will change based on the genus, so we shall focus on the genus zero piece. It would be interesting to work out the contributions for genus one to confirm that we have the same pattern there.

The idea of the Lunin-Mathur method is to lift this correlator onto the cover where the twist operators are ramified into identity operators. The fact that we have a non-zero central charge means that the contribution of the twist fields will be accounted for by a Liouville term. The various regulators that are introduced for this computation will then give non-zero contributions. A properly normalized four point function will remove the regulators, leaving the desired result. To do this, we need to first have a map with the correct ramification around each of the twist insertions. In this case, one choice is the map
\begin{equation}
z=Ct^n\frac{t-a}{t-1}.
\end{equation}
The ramifications of $z=0$ and $z=\infty$ are clear, the other two points are at
\begin{equation}
t_{\pm}=\frac{1}{2n}\left[(n-1)a+n+n\pm\sqrt{(a-1)((n-1)^2a-(n+1)^2)}\right]\,.
\end{equation}
These need to correspond to our two points on the base, $z=1$ and $z=w$. By choosing $z(t_+)=1$, we get a constraint on the undetermined coefficient
\begin{equation}
C=t_{+}^{-n}\frac{t_+-1}{t_+-a}\,,
\end{equation}
then we will also have $z(t_-)=w$.

The result of the procedure is
\begin{align}\label{4pfinCa}
&\frac{\langle\sigma_n(0,0)\sigma_2(1,1)\sigma_2(w,\bar{w})\sigma_n(\infty,\bar{\infty})\rangle}{\langle\sigma_2(0,0)\sigma_2(1,1)\rangle\langle\sigma_n(0,0)\sigma_n(\infty,\bar{\infty})\rangle}\\
&=|C|^{-\frac{c}{4}}|a|^{-\frac{c}{12}\left(\frac{n+1}{2}-\frac{1}{n}\right)}|1-a|^{-\frac{c}{8}}|(n-1)^2a-(n+1)^2|^{-\frac{c}{24}}n^{-\frac{c}{12}}2^{-\frac{5c}{12}}\,,\nonumber
\end{align}
which is a generalization of the result in equation (7.23) of \cite{Lunin:2000yv}. The authors had worked with $c=1$, whereas we have kept it general. With this in hand, we can now show how to extract the information of the OPE from the four point function.

\subsection{Taking the coincidence limit}

With the four point function, we can exploit the coincidence limit to constrain the operators that occur in the OPE of $\sigma_2(w,\bar{w})\sigma_n(0,0)$. This is done by taking the $(w,\bar{w})\rightarrow(0,0)$ limit of the four point function.

Consider the (carefully normalized) OPE of the two twist operators,
\begin{equation}
\frac{\sigma_2(w,\bar{w})\sigma_n(0,0)}{\langle\sigma_2(0,0)\sigma_2(1,1)\rangle^{1/2}\langle\sigma_n(0,0)\sigma_n(1,1)\rangle^{1/2}}=\sum\limits_k|w|^{-2h_2-2h_n}w^{h_k}\bar{w}^{\bar{h}_k}\,C_{2nk}\,\frac{\mathcal{O}_k(0,0)}{\langle\mathcal{O}_k|\mathcal{O}_k\rangle^{1/2}}\,.
\end{equation}
Here $\mathcal{O}_k$ are the primaries and descendant fields that show up in the expansion, and $C_{ijk}$ are the corresponding structure constants. We have also exploited the fact that $h_n=\bar{h}_n$ for the twist operators. Plugging this into the four point function, we can see that
\begin{align}\nonumber
&\frac{\langle\sigma_n(0,0)\sigma_2(1,1)\sigma_2(w,\bar{w})\sigma_n(\infty,\bar{\infty})\rangle}{\langle\sigma_2(0,0)\sigma_2(1,1)\rangle\langle\sigma_n(0,0)\sigma_n(\infty,\bar{\infty})\rangle}\\
&=\sum\limits_k|w|^{-2h_2-2h_n}w^{h_k}\bar{w}^{\bar{h}_k}\,C_{2nk}\,\frac{\langle\mathcal{O}_k(0,0)\sigma_2(1,1)\sigma_n(\infty,\bar{\infty})\rangle\langle\sigma_n(0,0)\sigma_n(1,1)\rangle^{1/2}}{\langle\sigma_2(0,0)\sigma_2(1,1)\rangle^{1/2}\langle\mathcal{O}_k|\mathcal{O}_k\rangle^{1/2}\langle\sigma_n(0,0)\sigma_n(\infty,\bar{\infty})\rangle}\\
&=\sum\limits_k|w|^{-2h_2-2h_n}w^{h_k}\bar{w}^{\bar{h}_k}\,C_{2nk}^2\,.
\end{align}
Therefore, the three point function has the proper normalization to return the structure constant. Since we have the full expression for the four point function in equation (\ref{4pfinCa}), we can expand this in powers of $w$ and obtain the data on the spectrum of $(h_k,\bar{h}_k,C_{2nk})$.

Taking the expansion of the four point function around $w=0$ is subtle as $w$ is a complicated function of $a$,
\begin{align}
w=&\left(\frac{n+1+(n-1)a-\sqrt{(a-1)((n-1)^2a-(n+1)^2)}}{n+1+(n-1)a+\sqrt{(a-1)((n-1)^2a-(n+1)^2)}}\right)^n\\\nonumber
&\times\frac{n+1-(n-1)a-\sqrt{(a-1)((n-1)^2a-(n+1)^2)}}{n+1-(n-1)a+\sqrt{(a-1)((n-1)^2a-(n+1)^2)}}\,,
\end{align}
which is in general not invertible. Instead, we can invert this expression order by order and use this expansion to expand the original four point function in powers of $w$. This will not give us information about all of the operators that are exchanged at arbitrary order, but we can still see some of the lowest lying exchanged operators. So we expand $w$ in terms of $a$, finding
\begin{align}
w&=\frac{n^{2n}}{(n+1)^{2n+2}}a^{n+1}+\frac{2n^{2n+1}}{(n+1)^{2n+3}}a^{n+2}+\frac{n^{2n+1}(3n^2+3n+1)}{(n+1)^{2n+5}}a^{n+3}\\\nonumber
&+\frac{2n^{2n+1}(1+5n+12n^2+12n^3+6n^4)}{3(n+1)^{2n+7}}a^{n+4}+\mathcal{O}(a^{n+5})\,.
\end{align}
We can invert this expansion to find $a$ in terms of $w$,
\begin{align}
a&=(n+1)^2n^{-\frac{2n}{n+1}}w^{\frac{1}{n+1}}-2(n+1)^2n^{\frac{1-3n}{n+1}}w^{\frac{2}{n+1}}-n^{\frac{1-5n}{n+1}}(n+1)^2(1-5n+n^2)w^{\frac{3}{n+1}}\\\nonumber
&-2(n+1)^2n^{\frac{1-7n}{n+1}}(1-10n+24n^2-10n^3+n^4)w^{\frac{4}{n+1}}/3+\mathcal{O}(w^{\frac{5}{n+1}})\,.
\end{align}
We can then plug this expression into equation (\ref{4pfinCa}) to find the expansion
\begin{align}\label{coincidentlimitexpansion}
\frac{\langle\sigma_n(0,0)\sigma_2(1,1)\sigma_2(w,\bar{w})\sigma_n(\infty,\bar{\infty})\rangle}{\langle\sigma_2(0,0)\sigma_2(1,1)\rangle\langle\sigma_n(0,0)\sigma_n(\infty,\bar{\infty})\rangle}&=|w|^{-\frac{c(n^2+n-2)}{24n(n+1)}}2^{-\frac{5c}{12}} n^{-\frac{c(2n^2+3n+3)}{12(n+1)}}(n+1)^{\frac{c(2n+n+2)}{12n}} \\ \nonumber
& \times \left(1+ \frac{c}{8}n^{\frac{2-2n}{n+1}}w^{\frac{2}{n+1}}+\frac{2c}{9}(n-1)^2n^{\frac{2-4n}{n+1}}w^{\frac{3}{n+1}}+...\right) \times \text{c.c.}\,,
\end{align}
where we have suppressed the complex conjugate of the brackets that come from the antiholomorphic sector. The leading term of this was already found and examined in \cite{Lunin:2000yv}.

This expression has a useful form. The power of the leading $w$ term is
\begin{equation}
h_{n+1}-h_2-h_n=\frac{c}{24}\left(n+1-\frac{1}{n+1}-2+\frac{1}{2}-n-\frac{1}{n}\right)=-\frac{c(n^2+n-2)}{48n(n+1)}\,,
\end{equation}
and the corresponding weight for the antiholomorphic sector. Then the powers in the brackets indicate operators that have weight
\begin{equation}
h_{n+1}+\frac{k}{n+1}\,.
\end{equation}
These powers represents the fractional excitations of the bare twist $\sigma_{n+1}$. In the next section, we will analyze this expansion and see how we can reconstruct it completely from the exchange of fractionally moded Virasoro excitations of the bare twist operator.

\section{Finding the exchanged operators}\label{exchangereconstruct}
\subsection{First non-leading order}

The first immediate result we can see is that there is no term in the series with an excitation of ${1}/{(n+1)}$. This is consistent with our results in section \ref{primaryfractionalmodes}, where we found that the $L_{-1/n}\sigma_n$ operator vanished. So even if we did not use the covering space method to show how this operator vanished, we can see it directly from the coincidence limit. There are no other operators that contribute at this order either, which is consistent with our conjecture that only fractional Virasoro excitations show up in the OPE.

\subsection{Second non-leading order}

This order is non-vanishing and so we should now consider the three point function to match the results. Specifically, we are interested in examining the normalized three point function
\begin{equation}
\frac{\langle L_{-\frac{2}{n+1}}\sigma_{n+1}(0,0)\sigma_2(1,1)\sigma_n(\infty,{\bar{\infty}})\rangle}{\langle\sigma_n(0,0)\sigma_n(\infty,{\bar{\infty}})\rangle}
{\frac{\langle\sigma_n(0,0)\sigma_n(1,1)\rangle^{1/2}}{\langle\sigma_2(0,0)\sigma_2(1,1)\rangle^{1/2}\langle L_{-\frac{2}{n+1}}\sigma_{n+1}(0,0)L_{-\frac{2}{n+1}}\sigma_{n+1}(1,1)\rangle^{1/2}}}\,,
\end{equation}
by lifting this to the cover using the map
\begin{equation}\label{3pfmap}
z(t)=\frac{t^{n+1}}{(n+1)t-n}\,.
\end{equation}

We first need to work out the results from the Liouville term. The formula (D.82) of \cite{Avery:2010qw} allows us to find the normalized contribution,
\begin{equation}
Z^{norm}=\left(\prod\limits^M_{i=1}p_i^{\frac{c}{12}(p_i+1)}\right)\left(\prod\limits^{N-1}_{j=0}q_j^{\frac{c}{12}(q_j-1)}\right)\left(\prod\limits^M_{i=1}|a_i|^{-\frac{c}{12}\frac{p_i-1}{p_i}}\right)\left(\prod\limits^F_{j=0}|b_j|^{-\frac{c}{12}\frac{q_j+1}{q_j}}\right)|b_0|^{\frac{c}{6}}q_0^{\frac{c}{6}}\,.
\end{equation}
where we have dropped the infinity regulator, as we have carefully normalized the bare twists correctly with the appropriate two point functions. We find that for our three point function and map, $M=2$, $N=1$, $F=1$, $p_1=n+1$, $p_2=2$, $q_0=n$, $q_1=1$, $a_1=-{1}/{n}$, $a_2=n(n+1)/2$, $b_0=1/(n+1)$, and $b_1=n^{n+1}/(n+1)^{n+2}$. Putting it all together, we find
\begin{equation}\label{3pfbaretwist}
Z^{norm}=2^{-\frac{5c}{24}}(n+1)^{\frac{c(2+n+2n^2)}{24n}}n^{-\frac{c(3+3n+2n^2)}{24(n+1)}}\,.
\end{equation}

We can now lift the operator up to the cover to find the contribution from the excitation,
\begin{align}
L_{-2/(n+1)}\sigma_{n+1}&\rightarrow\oint\frac{dt}{2\pi i}z(t)^{\frac{n-1}{n+1}}\left(\frac{dz}{dt}\right)^{-1}\left(T(t)-\frac{c}{12}\{z,t\}\right)\\
&=\left(\frac{n^{\frac{2}{n+1}}}{n+1}T(0)-\frac{cn^{\frac{1-n}{n+1}}}{4(n+1)}\right)\,.
\end{align}
This is obtained by expanding in powers of $1/t$ and computing the required residue integral. There is also an extra phase here, coming from the fact that the expansion of the map has a negative coefficient. Since this phase cancels out in the end, we will suppress it.

Since the one point function vanishes, it is only the Schwarzian that contributes. We see the contribution from the non-twist is then
\begin{equation}\label{Lm2dn1NonTwistResult}
\langle L_{-2/(n+1)}\sigma_{n+1}(0,0)\sigma_2(1,1)\sigma_n(\infty,{\bar{\infty}})\rangle_{\rm non-twist}=\frac{cn^{\frac{1-n}{n+1}}}{4(n+1)}\,.
\end{equation}
Finally, we need to have the normalization. This too can be obtained by lifting to the cover. Equivalently, we can make use of equation (\ref{normformkdn}) with $k=2$ (and replacing $n$ with $n+1$), which gives the result
\begin{equation}
\langle L_{-2/(n+1)}\sigma_{n+1}(0,0)\, L_{-2/(n+1)}\sigma_{n+1}(1,1)\rangle=\frac{c}{2(n+1)^2}\,.
\end{equation}
We can put this together to find the normalized three point function,
\begin{align}\nonumber
&\frac{\langle L_{-\frac{2}{n+1}}\sigma_{n+1}(0,0)\sigma_2(1,1)\sigma_n(\infty,{\bar{\infty}})\rangle}{\langle\sigma_n(0,0)\sigma_n(\infty,{\bar{\infty}})\rangle}
{\frac{\langle\sigma_n(0,0)\sigma_n(1,1)\rangle^{1/2}}{\langle\sigma_2(0,0)\sigma_2(1,1)\rangle^{1/2}\langle L_{-\frac{2}{n+1}}\sigma_{n+1}(0,0)L_{-\frac{2}{n+1}}\sigma_{n+1}(1,1)\rangle^{1/2}}}\\
&=(2^{-\frac{5c}{24}}(n+1)^{\frac{c(2+n+2n^2)}{24n}}n^{-\frac{c(3+3n+2n^2)}{24(n+1)}})\left(\frac{cn^{\frac{1-n}{n+1}}}{4(n+1)}\right)\left(\frac{c}{2(n+1)^2}\right)^{-1/2}\\
&=\sqrt{\frac{c}{8}}n^{\frac{1-n}{n+1}}\left(2^{-\frac{5c}{24}}(n+1)^{\frac{c(2+n+2n^2)}{24n}}n^{-\frac{c(3+3n+2n^2)}{24(n+1)}}\right)\,.\label{Lm2dn1Result}
\end{align}

To compare to the coefficient of the $|w|^{-\frac{c(n^2+n-2)}{12n(n+1)}}w^{\frac{2}{n+1}}$ in equation (\ref{coincidentlimitexpansion}), we need to square this result. We find
\begin{equation}
\frac{c}{8}n^{\frac{2-2n}{n+1}}\left(2^{-\frac{5c}{12}}(n+1)^{\frac{c(2+n+2n^2)}{12n}}n^{-\frac{c(3+3n+2n^2)}{12(n+1)}}\right)\,,
\end{equation}
which matches the coefficient of the coincidence limit. This indicates the that operator $L_{-2/(n+1)}\,\sigma_{n+1}$ fully accounts for the exchange at this order.

\subsection{Third non-leading order}

The third non-leading order is also non-vanishing. So we can examine the three point function
\begin{equation}
\frac{\langle L_{-\frac{3}{n+1}}\sigma_{n+1}(0,0)\sigma_2(1,1)\sigma_n(\infty,{\bar{\infty}})\rangle}{\langle\sigma_n(0,0)\sigma_n(\infty,{\bar{\infty}})\rangle}
{\frac{\langle\sigma_n(0,0)\sigma_n(1,1)\rangle^{1/2}}{\langle\sigma_2(0,0)\sigma_2(1,1)\rangle^{1/2}\langle L_{-\frac{3}{n+1}}\sigma_{n+1}|L_{-\frac{3}{n+1}}\sigma_{n+1}\rangle^{1/2}}}\,.
\end{equation}
This is an interesting case, as for $n=2$ the operator is simply $L_{-1}$, i.e., a descendant of the $\sigma_3$ operator. This contribution is fixed by conformal invariance. We can proceed as we did before, lifting to the cover,
\begin{align}
L_{-3/(n+1)}\sigma_{n+1}&\rightarrow\oint\frac{dt}{2\pi i}z(t)^{\frac{n-2}{n+1}}\left(\frac{dz}{dt}\right)^{-1}\left(T(t)-\frac{c}{12}\{z,t\}\right),\\
&=\oint\frac{dt}{2\pi i}\left(\frac{((n+1)t-n)^{\frac{n+4}{n+1}}}{n(n+1)t^3(t-1)}T(t)
\right.
\\\nonumber
&\left. +\frac{c((n+1)t-n)^{\frac{n+4}{n+1}}((n^2-1)t^2-(2n^2+2n-4)t+n(n+2))}{24n(n+1)t^4(t-1)^3}\right)\,.
\end{align}
Again, we will not need the $T(t)$ term, so we focus on just the Schwarzian term. Doing this integral gives
\begin{equation}\label{nonnorm3pfdescendant}
\langle L_{-3/(n+1)}\sigma_{n+1}(0,0)\sigma_2(1,1)\sigma_n(\infty,{\bar{\infty}})\rangle_{\rm non-twist}=\frac{2c(n-1)}{3(n+1)n^{\frac{2n-1}{n+1}}}\,.
\end{equation}
To compute the normalization, we have to be careful to note that for the case of $n=2$, this operator is a descendant. So normalizing with the two point function at finite points will not get the normalization correct. Instead, we will use equation (\ref{normformkdn}) with $k=3$ (and replacing $n$ with $n+1$) to find the normalization
\begin{equation}
\langle L_{-3/(n+1)}\sigma_{n+1}|L_{-3/(n+1)}\sigma_{n+1}\rangle=\frac{2c}{(n+1)^2}\,.
\end{equation}

Putting these two contributions together with the Liouville term in equation (\ref{3pfbaretwist}), we find
\begin{align}\nonumber
&\frac{\langle L_{-\frac{3}{n+1}}\sigma_{n+1}(0,0)\sigma_2(1,1)\sigma_n(\infty,{\bar{\infty}})\rangle}{\langle\sigma_n(0,0)\sigma_n(\infty,{\bar{\infty}})\rangle}
{\frac{\langle\sigma_n(0,0)\sigma_n(1,1)\rangle^{1/2}}{\langle\sigma_2(0,0)\sigma_2(1,1)\rangle^{1/2}\langle L_{-\frac{3}{n+1}}\sigma_{n+1}|L_{-\frac{3}{n+1}}\sigma_{n+1}\rangle^{1/2}}}\\
&=\left(2^{-\frac{5c}{24}}(n+1)^{\frac{c(2+n+2n^2)}{24n}}n^{-\frac{c(3+3n+2n^2)}{24(n+1)}}\right)\left(\frac{2c(n-1)}{3(n+1)n^{\frac{2n-1}{n+1}}}\right)\left(\frac{2c}{(n+1)^2}\right)^{-1/2}\\
&=\sqrt{\frac{2c}{9}}n^{\frac{1-2n}{n+1}}(n-1)\left(2^{-\frac{5c}{24}}(n+1)^{\frac{c(2+n+2n^2)}{24n}}n^{-\frac{c(3+3n+2n^2)}{24(n+1)}}\right)\,.
\end{align}
To compare to the coefficient of the $|w|^{-\frac{c(n^2+n-2)}{12n(n+1)}}w^{\frac{3}{n+1}}$ in equation (\ref{coincidentlimitexpansion}), we need to square this result. We find
\begin{equation}
\frac{2c}{9}n^{\frac{2-4n}{n+1}}(n-1)^2\left(2^{-\frac{5c}{12}}(n+1)^{\frac{c(2+n+2n^2)}{12n}}n^{-\frac{c(3+3n+2n^2)}{12(n+1)}}\right)\,,
\end{equation}
which matches the coefficient of the coincidence limit. This indicates that the operator $L_{-3/(n+1)}\,\sigma_{n+1}$ fully accounts for the exchange at this order. This is also true when $n=2$, where this term is a descendant and the result is fixed using conformal invariance. We have done the computation with this method as well, finding agreement.

\subsection{Higher orders}

So far, we have been exploring the cases of operators with $L_{-k/(n+1)}\,\sigma_{n+1}$ with $k\leq n+1$. The next obvious step would be to examine operators with $k>n+1$. As we discussed in section \ref{primaryfractionalmodes}, these operators will not be primary on their own, and we must project out the non-primary parts. In addition, we will have to consider the fact that the descendants of lower weight operators will also contribute in the coincidence limit. Furthermore, as we consider higher and higher weights, the number of possible operators will increase, making it more difficult to do the matching for general $n$.

To explore operators with $k>n+1$ with the coincidence limit, we will investigate the simple non-trivial case with $n=2$. We therefore look at the operator $L_{-5/3}\,\sigma_3$, with its proper projection. Substituting $n=3$ into equation (\ref{Lmn3dn1}), we find that
\begin{equation}
\left(L_{-5/3}-\frac{12}{c+6}L_{-1}L_{-2/3}\right)\sigma_3
\end{equation}
is the primary part of $L_{-5/3}\,\sigma_3$. Then we would have this operator possibly contributing, along with the descendant $L_{-1}L_{-2/(n+1)}\,\sigma_{n+1}$. This projection procedure ensures these operators are orthogonal, so they will contribute without any overlap.

We can compute the three point function,
\begin{equation}
\langle \left(L_{-5/3}-\frac{12}{c+6}L_{-1}L_{-2/3}\right)\sigma_{3}(0,0)\sigma_2(1,1)\sigma_2(\infty,{\bar{\infty}})\rangle\,.
\end{equation}
This can be separated into two pieces. Let us first examine
\begin{equation}
\langle L_{-5/3}\sigma_{3}(0,0)\sigma_2(1,1)\sigma_2(\infty,{\bar{\infty}})\rangle\,.
\end{equation}
To do this, we lift to the cover using the map in equation (\ref{3pfmap}), finding
\begin{align}
L_{-5/3}\,\sigma_{3}(0,0)&\rightarrow\oint\frac{dt}{2\pi i}z(t)^{-2/3}\left(\frac{dz}{dt}\right)^{-1}\left(T(t)-\frac{c}{24}\{z,t\}\right)\\
&=\frac{2^{-1/3}c}{9}\,.
\end{align}
Note we can have dropped the $T(t)$ dependent terms, as they will not contribute to the correlator. This also makes the residue integral much easier to compute. So the correlator excluding the Liouville contribution is
\begin{equation}
\langle L_{-5/3}\,\sigma_{3}(0,0)\sigma_2(1,1)\sigma_2(\infty,{\bar{\infty}})\rangle_{\rm non-twist}=\frac{2^{-1/3}c}{9}\,.
\end{equation}
Now we can compute the second part,
\begin{equation}
\langle L_{-1}L_{-2/3}\,\sigma_{3}(0,0)\sigma_2(1,1)\sigma_2(\infty,{\bar{\infty}})\rangle\,.
\end{equation}
We can compute this by lifting to the cover, or by using the fact that $L_{-1}$ acts as derivative through the Ward identity. Both methods agree; here we will show the Ward identity method. We find
\begin{equation}
\langle L_{-1}L_{-2/3}\,\sigma_{3}(0,0)\sigma_2(1,1)\sigma_2(\infty,{\bar{\infty}})\rangle=h_{L_{-2/3}\,\sigma_3}\langle L_{-2/3}\,\sigma_{3}(0,0)\sigma_2(1,1)\sigma_2(\infty,{\bar{\infty}})\rangle\,.
\end{equation}
This is the only term that survives when we properly take the limit of the third operator going to infinity. The weight of the first operator, $h_{L_{-2/3}\sigma_3}$, can be found simply by adding the weight of the bare twist together with the excitation
\begin{equation}
h_{L_{-2/3}\sigma_3}=\frac{c}{24}(3-\frac{1}{3})+\frac{2}{3}=\frac{c+6}{9}\,.
\end{equation}
Now we can include the result from the lift done in equation (\ref{Lm2dn1NonTwistResult}) with $n=2$. This will result in
\begin{equation}
\langle L_{-1}L_{-2/3}\,\sigma_{3}(0,0)\sigma_2(1,1)\sigma_2(\infty,{\bar{\infty}})\rangle_{\rm non-twist}=\frac{c+6}{9}\times \frac{c}{2^{7/3}3}=\frac{c(c+6)}{2^{7/3}3^{3}}\,.
\end{equation}
Putting the two pieces of the full three point function together, we find
\begin{equation}
\langle \left(L_{-5/3}-\frac{12}{c+6}L_{-1}L_{-2/3}\right)\sigma_{3}(0,0)\sigma_2(1,1)\sigma_2(\infty,{\bar{\infty}})\rangle_{non-twist}=\frac{2^{-1/3}c}{9}-\frac{12}{c+6}\frac{c(c+6)}{2^{7/3}3^{3}}=0\,,
\end{equation}
which is an interesting result.

To double check this, we show that the contribution of the descendant $L_{-1}L_{-2/3}\sigma_3$ constitutes the full piece to the four point function. We first need the normalization, which can be obtained via the fractional Virasoro algebra and results in
\begin{equation}
\langle L_{-1}L_{-2/3}\,\sigma_3|L_{-1}L_{-2/3}\,\sigma_3\rangle=\frac{c(c+6)}{81}\,.
\end{equation}
Then the full normalized result (including the Liouville term) is then
\begin{align}
\frac{\langle L_{-1}L_{-2/3}\,\sigma_{3}(0,0)\sigma_2(1,1)\sigma_2(\infty,{\bar{\infty}})\rangle}{\langle L_{-1}L_{-2/3}\,\sigma_3|L_{-1}L_{-2/3}\,\sigma_3\rangle^{1/2}}&=\left(\frac{c(c+6)}{2^{7/3}3^{3}}\times2^{-4c/9}3^{c/4}\right)\left(\frac{c(c+6)}{81}\right)^{-1/2}\\
&=\sqrt{c(c+6)}2^{-7/3-4c/9}3^{c/4-1}\,.\label{Lm1Lm2d3Result}
\end{align}
We then need to expand out to the correct order. In this case we are interested in
\begin{equation}
w^{h_{L_{-5/3}\sigma_3}-2h_{\sigma_2}}\bar{w}^{\bar{h}_{\sigma_3}-2\bar{h}_{\sigma_2}}=|w|^{-c/36}w^{5/3}\,.
\end{equation}
We can carry this out by expanding equation (\ref{4pfinCa}) to this order, finding
\begin{align}
&\frac{\langle\sigma_2(0,0)\sigma_2(1,1)\sigma_2(w,\bar{w})\sigma_2(\infty,\bar{\infty})\rangle}{\langle\sigma_2(0,0)\sigma_2(1,1)\rangle\langle\sigma_2(0,0)\sigma_2(\infty,\bar{\infty})\rangle}=|w|^{-c/36}2^{-8c/9}3^{c/2}\\\nonumber
&\times(1+\frac{c}{2^{11/3}}w^{2/3}+\frac{c}{18}w+\frac{c(c+6)}{2^{25/3}}w^{4/3}+\frac{c(c+6)}{9\times2^{14/3}}w^{5/3}+\frac{c(9000+998c+27c^2)}{331776}w^2)\times\text{c.c.}\,.
\end{align}
We can see that the coefficient is then
\begin{equation}
2^{-8c/9}3^{c/2}\frac{c(c+6)}{9\times2^{14/3}}=(\sqrt{c(c+6)}2^{-7/3-4c/9}3^{c/4-1})^2\,,
\end{equation}
where the argument in the square is the result we found in equation (\ref{Lm1Lm2d3Result}).

This shows that there is no contribution from the primary to the four point function: the latter is entirely accounted for by the descendant $L_{-1}L_{-2/3}\,\sigma_3$. We may suspect that the lack of a contribution may mean that $(L_{-5/3}-L_{-1}L_{-2/3})$ is a null state, like $L_{-1/(n+1)}\sigma_{n+1}$. But recruiting the fractional Virasoro algebra shows us that
\begin{equation}
\langle \left(L_{-5/3}-\frac{12}{c+6}L_{-1}L_{-2/3}\right)\sigma_{3}\,|\left(L_{-5/3}-\frac{12}{c+6}L_{-1}L_{-2/3}\right)\sigma_{3}\rangle=\frac{c(10c+44)}{9(c+6)}\,,
\end{equation}
which is not zero and hence it is not a null operator. Note that this vanishing result does not generalize to higher twist. We can compute the general correlator
\begin{equation}
\langle \left(L_{-(n+3)/(n+1)}-\frac{24(n+2)}{c((n+1)^2-1)+48}L_{-1}L_{-2/(n+1)}\right)\sigma_{n+1}(0,0)\sigma_2(1,1)\sigma_n(\infty,\bar{\infty}) \rangle\,,
\end{equation}
and it will, in general, be a complicated non-zero function of $n$. It would be interesting to understand why this happens to vanish for $n=2$.

\section{Discussion}\label{discussion}

With the work in this paper, we have explored the OPE of twist operators in a general symmetric orbifold ($M^N/S_N$) CFT. We studied the fractional Virasoro modes, which are exclusive to the twisted sector owing to the twisted boundary conditions. We have shown how the $L_{-1/n}\sigma_n$ operator vanishes, as well as how the fractionally moded Virasoro modes form a generalization of the regular Virasoro algebra.  To explore how these operators show up in the OPE of twist operators, we examined the coincidence limit of the four point function $\langle \sigma_n\sigma_2\sigma_2\sigma_n\rangle$, which gave us details on the $\sigma_n\times\sigma_2$ OPE.  We further showed the vanishing of the term corresponding to $L_{-1/n}\sigma_n$ in the coincidence limit, as well as matching both primary and descendant terms to several non-leading powers in the expansion. 
While our calculations only checked the first few orders, we conjecture that this pattern will persist to all orders in the OPE: operators constructed using the bare twist and fractional Virasoro modes will be the only operators that show up in the OPE of bare twist operators for all symmetric orbifold CFTs. 

From the point of view of the covering space, these results are not very surprising. When we lift bare twists to the cover, they are replaced by the identity. Then the OPE of the identity should contain the identity, plus all the descendants under the Virasoro algebra on the cover. These descendants are the covering space representatives of the fractional Virasoro excitations. This would also explain the vanishing of $L_{-1/n}|\sigma_n\rangle$, as this would roughly lift to $L_{-1}|0\rangle$, which vanishes.

Given that the OPEs and three point functions appear to depend on an underlying algebra, and that this underlying algebra is essentially the Virasoro algebra, we may be tempted to find the OPE using more algebraic techniques.  In a CFT, if we know the coefficient of primary $\phi_p$ appearing in the OPE of two other primaries $\phi_1$ and $\phi_2$, then the descendants of $\phi_p$ must also appear with coefficients that are given by conformal invariance \cite{DiFrancesco:1997nk}.  Explicitly, the OPE is given by
\begin{equation}
\phi_1(z) \phi_2(0)=\sum_p \sum_{k,\bar{k}} \,C_{12}^p \,\beta_{12}^{p \{k\}} \,\bar{\beta}_{12}^{p \{\bar{k}\}} \,z^{h_p-h_1-h_2+k} \bar{z}^{\bar{h}_p-\bar{h}_1-\bar{h}_2+{\bar{k}}} \,\phi_p^{\{k,\bar{k}\}}\,,
\end{equation}
where the $\beta_{12}^{p \{k\}}$ and $\bar{\beta}_{12}^{p \{\bar{k}\}}$ are a set of coefficients, $\{k\}$ and $\{\bar{k}\}$ are a list of negative mode Virasoro generators used to excite $\phi_p$.  The coefficients $\beta$ and $\bar{\beta}$ are read off by applying an arbitrary $L_n$ operator to both sides of the equation, arriving at a recurrence relation for $\beta$ and $\bar{\beta}$.  These recurrence relations can be solved iteratively to find the the coefficients $\beta$ and $\bar{\beta}$ level by level.

\begin{figure}
\begin{center}
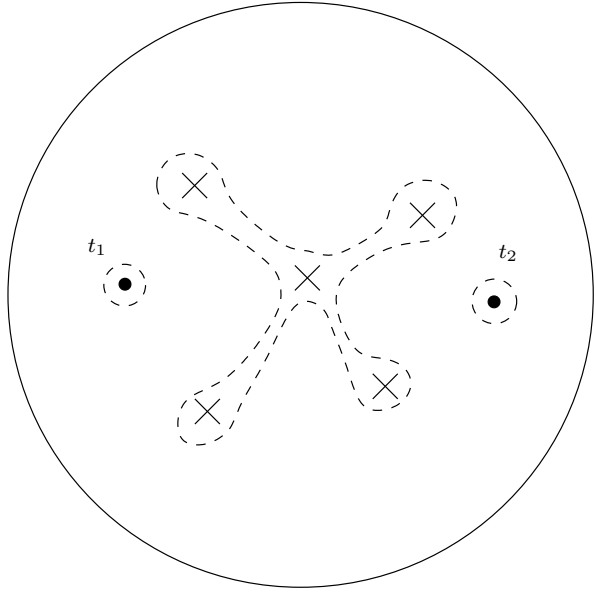
\end{center}
\caption{The diagram of the contour pull in the $t$ plane, attempting to pull the contour from infinity (solid line) to the other ramified points $t_1$ and $t_2$ (dashed line).  However, there is an additional obstruction coming from the branch points in $(z(t))^{k/(n+1)+1}$ (images of $z=0$), denoted by $\times$ in the diagram, leading to the contour being deformed around these points and accompanying branch cuts as well.  Depicted above is the case $n=4$.}
\label{fig1}
\end{figure}

In the previous sections, we have given evidence that there is a similar form for the OPE between two bare twists, except that the descendants are given by a list of fractional Virasoro modes.  Is it possible to get a recurrence relation for the $\beta$ coefficients in this case?  Let us take as a concrete example the fusion of two operators, say a $\sigma_2$ and a $\sigma_n$ fusing to a possibly excited $\sigma_{n+1}$.  Thus, we are concerned with the part of the OPE that reads
\begin{equation}
\sigma_2(z+a) \sigma_n(+a)= \cdots \sum_{k,\bar{k}} C_{2,n}^{n+1} \beta_{2,n}^{(n+1),\{k\}} \bar{\beta}_{2,n}^{(n+1) \{\bar{k}\}} z^{h_p-h_1-h_2+k} \bar{z}^{\bar{h}_p-\bar{h}_1-\bar{h}_2+{\bar{k}}} \sigma_{n+1}^{\{k,\bar{k}\}}(a)\,, \label{twistopeform}
\end{equation}
where the $\{k\}$ and $\{\bar{k}\}$ are a list of fractional modes on the right hand side, and we have shifted the two $\sigma$ operators by $a$, but then relax to $a\rightarrow 0$ at the end (for later convenience).  Applying a $L_{k/(n+1)}$ on the right hand side of (\ref{twistopeform}) appears to give no problem with $a=0$: the modified algebra should result in a simple way to read off the result.  However, on the left hand side of (\ref{twistopeform}), we must compute
\begin{equation}
L_{k/(n+1)} \,\sigma_{2}(z_1+a) \,\sigma_{n}(a)\,.
\end{equation}
This computation lifts to
\begin{equation}
\rightarrow \oint \frac{dt}{2\pi i}\left(\frac{dz}{dt}\right)^{-1}(z(t))^{k/(n+1)+1}\left(T(z)-\frac{c}{12}\left\{z(t),t\right\}\right)\,,
\end{equation}
where the map $z(t)$ has ramified points at $z(t=\infty)=\infty$, $z(t=t_1)=z_1+a$ and $z(t=t_2)=a$, and the contour in the $t$ plane is sufficiently large to go around the ramified points at $t=t_1$ and $t=t_2$.  The contour integral is well defined because $z\propto t^{n+1}$ as $t\rightarrow \infty$.  We want to deform this contour inward so that we may evaluate the left hand side of (\ref{twistopeform}). However, there are points that obstruct the deformation.  These are the singularities in the Schwarzian, occurring at $t=t_1$ and $t=t_2$, and the branch points in $(z(t))^{k/(n+1)+1}$, which occur at $z(t)=0$; there are $n+1$ such points.  Thus, the contour deforms as depicted in figure {\ref{fig1}}.

Thus, although the contour is single valued around $t=\infty$, as well as $t_1$ and $t_2$, the points where $z(t)=0$ lead to branch points in $(z(t))^{k/(n+1)+1}$, and so do not allow explicit computation.  Now it becomes clear why we used the regulator $a$: in the limit $a\rightarrow 0$ that the location of a ramified point and some of the images of $z=0$ coincide, making this point both ramified and a branch point (of order $n$).  We introduced $a$ as a way to separate these two distinct concerns.  In the case where $k$ is proportional to $n+1$, the contour pull may be performed, but this is just the case of the non-fractional Virasoro modes.  Thus, in the case of fractional modes, the computation is difficult to make sense of, and so does not lead to a recurrence relation for the coefficients $\beta$ and $\bar{\beta}$.

We can see a similar failure for the three point functions. In a conformal theory, the three point function between three primaries immediately leads to the form of the three point function between two of the primaries and a given descendant of the third.  The form of the correlator is written in terms of some differential operator acting on the original three point function of primaries. This computation again relies on contour pulls, and so will fail when trying to generalize to the fractional modes of the stress tensor in exactly the same way.  These two difficulties seem to be complementary, and cancel when exploring exchange channels in four point functions.  The four point function may be thought of as a combination of the OPE (with $\beta$ difficult to compute from contour pulls) and the three point function of fractional descendants (also difficult to compute with contour pulls).  Thus, one must always resort to other techniques, for example the covering space technique used here, to find the three point functions of fractional mode descendant operators.  There does not seem to be a straightforward algebraic approach to finding the OPE, or the three point functions, although this point warrants further investigation.

There are a number of open questions remaining. One obvious generalization is to consider supersymmetric orbifold CFTs. The extended SUSY algebra there offers a larger variety of possible structures to explore but we would still expect that the operators would only depend on the general properties of the permutation orbifold CFTs. We would also be interested to connect this back to computing the anomalous dimensions of low-lying non-protected operators in the D1D5 CFT \cite{Burrington:2012yq,Burrington:2017jhh}. It would also be interesting to understand this fractional Virasoro algebra more. We found that one of the primary fractional Virasoro excitations does not show up in the OPE. It would be interesting to see if this is part of a larger pattern and what that pattern might be. Studying these patterns may require connecting to the higher spin algebra found for the D1D5 CFT \cite{Gaberdiel:2014cha,Gaberdiel:2015mra,Gaberdiel:2015uca,Gaberdiel:2015wpo,Gaberdiel:2018rqv}.

\section*{Acknowledgements}
The work of AWP and ITJ is supported by a Discovery Grant from the Natural Sciences and Engineering Research Council of Canada. The work of BAB is supported by funds provided by Hofstra University, including a Faculty Research and Development Grant, and faculty startup funds.


\end{document}

%% file: ContourPull.pdftex_t
\begin{picture}(0,0)%
\includegraphics{ContourPull.pdf}%
\end{picture}%
%
%
\setlength{\unitlength}{3947sp}%
\begingroup\makeatletter\ifx\SetFigFont\undefined%
\gdef\SetFigFont#1#2#3#4#5{%
  \reset@font\fontsize{#1}{#2pt}%
  \fontfamily{#3}\fontseries{#4}\fontshape{#5}%
  \selectfont}%
\fi\endgroup%
\begin{picture}(3692,3690)(538,-2969)
\put(1047,-852){\makebox(0,0)[lb]{\smash{{\SetFigFont{8}{9.6}{\rmdefault}{\mddefault}{\updefault}{\color[rgb]{0,0,0}$t_1$}%
}}}}
\put(3629,-895){\makebox(0,0)[lb]{\smash{{\SetFigFont{8}{9.6}{\rmdefault}{\mddefault}{\updefault}{\color[rgb]{0,0,0}$t_2$}%
}}}}
\end{picture}%